\documentclass[graybox]{svmult}

\usepackage{mathptmx}       
\usepackage{helvet}         
\usepackage{courier}        
\usepackage{type1cm}        
\usepackage{makeidx}         
\usepackage{graphicx}        
\usepackage{multicol}        
\usepackage[bottom]{footmisc}
\usepackage{tikz}
\usepackage{xcolor}
\usetikzlibrary{matrix,shapes,arrows,positioning,chains}
\usepackage{hyperref}

\makeindex             

\begin{document}

\title*{A strategy for the matching of mobile phone signals with census data}

\author{Rodolfo Metulini and Maurizio Carpita}
\institute{
Rodolfo Metulini \at Department of Economic and Management, University of Brescia, Contrada Santa Chiara, 50,
	25122 Brescia, \email{rodolfo.metulini@unibs.it}
\and Maurizio Carpita \at Department of Economic and Management, University of Brescia, Contrada Santa Chiara, 50, 25122 Brescia, \email{maurizio.carpita@unibs.it}
\and \textbf{Please cite as:} Metulini, R., Carpita, M. (2019) A strategy for the matching of mobile phone signals with census data. SIS 2019 - Smart Statistics for Smart Applications - Book of short papers, editors: Giuseppe Arbia, Stefano Peluso, Alessia Pini, Giulia Rivellini. ISBN 9788891915108.
}
%
%
\maketitle

\abstract{Administrative data allows us to count for the number of residents.
The geo-localization of people by mobile phone, by quantifying the number of people at a given moment in time, enriches the amount of useful information for ``smart" (cities) evaluations.  
However, using Telecom Italia Mobile (TIM) data, we are able to characterize the spatio-temporal dynamic of the presences in the city of just TIM users.
A strategy to estimate total presences is needed. 
In this paper we propose a strategy to extrapolate the number of total people by using TIM data only. 
To do so, we apply a spatial record linkage of mobile phone data with administrative archives using the number of residents at the level of ``sezione di censimento".}

\section{Introduction}
\label{sec:intro}
A crucial aspect for the well-being of an urban area in the context of the much discussed ``Smart-City" argument refers to the monitoring of the dynamic of people' presences.
To do so, thanks to a research collaboration of DMS StatLab (\url{https://sites.google.com/a/unibs.it/dms-statlab/}) with the Statistical Office of the Municipality of Brescia, we use mobile phone data provided by Telecom Italia Mobile (TIM), quantifying the number of TIM users connected to the smartphone over a spatial grid characterized by its latitude and longitude, and over the time.

We estimate the presence of TIM users in the city of Brescia by classifying similar days in terms of both the spatial and the temporal dimension. 
In doing so, we use a larger ($\approx$ 2 years) dataset compared to that used in similar works (Zanini et al. \cite{zanini16understanding} for example).
To manage with high dimensional data, we employ a multi-stage procedure that, in the first step, converts the data matrix containing the values of the grid (2-D) at each point in time to a vector (1-D) of features using a method borrowed from image processing (the Histogram of Oriented Grandients - HOG).
The procedure, by classifying similar days using a mix of traditional (k-means) and functional data clustering techniques, is able to identify the dynamic related to the presences of TIM users in the city.
 
However, for a proper evaluation, all the people should be counted in.
In absence of the data about other companies' users, the coefficient related to TIM market share could be applied.
Unfortunately, this percentage is just available at country level, while we have reason to think that the market share varies along different cities due to socio-economic and demographic reasons.
So, the aim of this work focuses on the estimation of the number of people at a city level. 
In doing so, we employ a record linkage based on comparing residents from administrative archives with the number of TIM users on specific areas of the grid during a sample of proper time periods.

All in all, this work regards the match of administrative archives with data from technological devices, adopting a method that reminds spatial record linkages (Blum and Calvo \cite{blum01geo} and Xu et al. \cite{xu12geo}), for the purpose of quantifying the presences in the city when some data are missings. 

Section \ref{sec:data} describes the mobile phone data, section \ref{sec:proc} summarizes the procedure adopted to classify similar days and to quantify the number of TIM users in those days. 
Section \ref{sec:strat} outlines the record linkage strategy adopted and section  \ref{sec:conc} discusses and concludes.   
\section{Data}
\label{sec:data}
This work focuses on mobile phone data provided by Telecom Italia Mobile (TIM), which is currently the largest operator in Italy in this sector. 
These data sometimes goes with the name of ``Erlang" measures.
In detail, the data refer to the mobile phone activity recorded in the period April 1st 2014 to August 11th 2016, in a rectangular region defined by latitude 45.21$^{\circ}$ N - 46.36$^{\circ}$  N and longitude 9.83$^{\circ}$  N - 10.85$^{\circ}$  N. 
Data were aggregated into 923 x 607 cells of 150 $m^2$ size each. 
Data are available at intervals of 15 minutes, for a total of more than 40,000 millions of records collected. 
For each cell and for each time interval, the corresponding record refers to the average number of mobile phones simultaneously connected to the network in that area in that time interval.
The mobility feature of these data is hidden, in the sense it is not possible to trace the single person over the time. \footnote{Works following the single person over the time exist. Unfortunately, those kind of data are available subject to the administration of a survey, that is usually limited in terms of time and usually cover a small portion of the population (Zhaedi and Shafahi \cite{zahedi18estimating}).}
Similar data has been used by Carpita and Simonetto \cite{carpita14big}, who analyzed the presence of people during big events in the city of Brescia, by Zanini et al. \cite{zanini16understanding}, who find, by mean of a Independent Component Analysis (ICA), a number of spatial components that separate main areas of the city of Milano, and by others (Metulini and Carpita \cite{metulini18on} Manfredini et al. \cite{manfredini15treelet} and Secchi et al. \cite{secchi17analysis}).
\section{Estimating TIM Density Profiles}
\label{sec:proc}
\begin{figure}[htbp!]
	\centering
\includegraphics[scale=0.32]{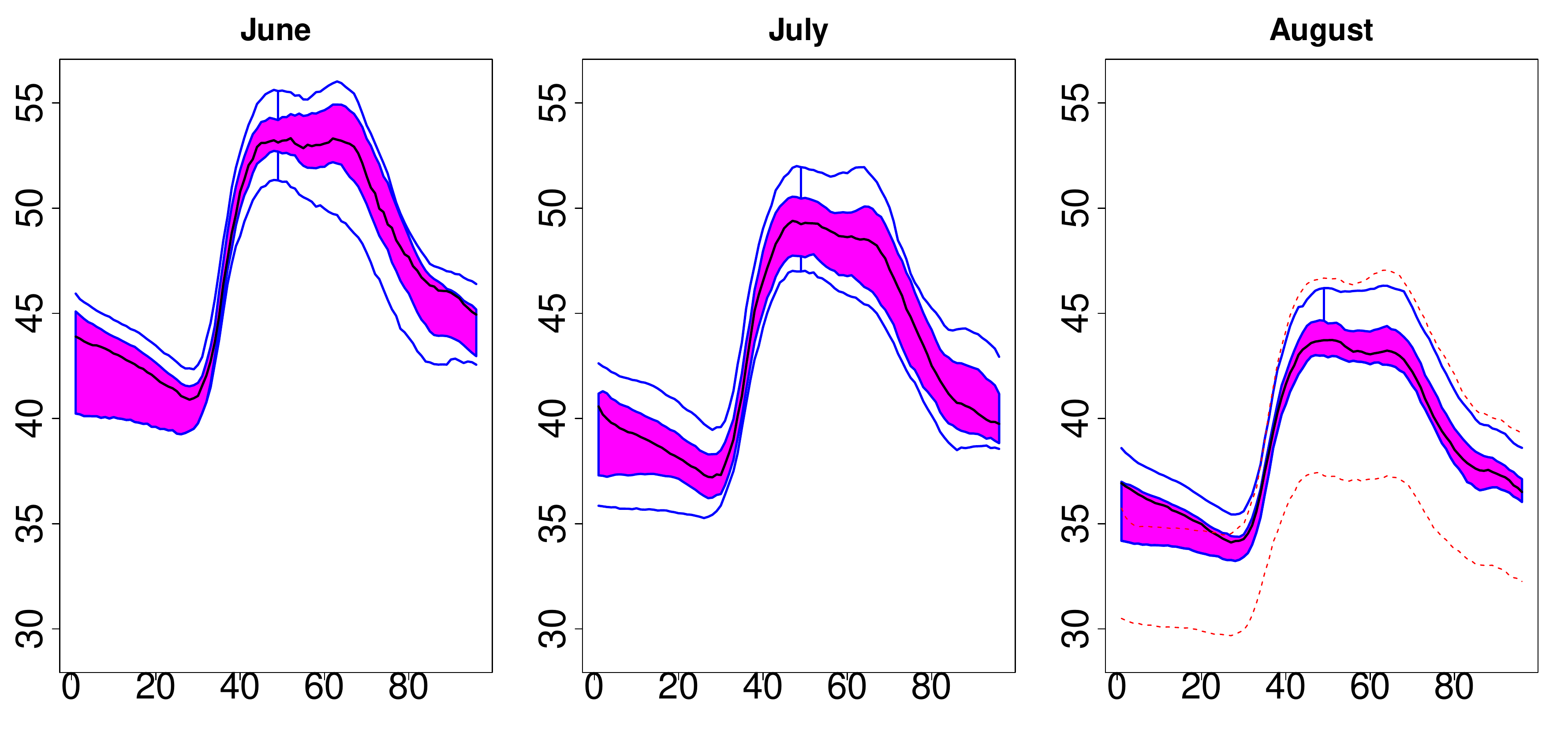}
	\caption{Functional box plots of the Daily Density Profiles of Summer 2016 (y-axis, in thousands of people), divided by month. Quarters are in the x-axis. Median (black curve), envelope (purple area), 1.5 * envelope (blue curves), outliers (red curves). Our Elaborations.}
	\label{fig:bplot}
\end{figure}

To characterize the dynamic of the presences in the city by defining daily density profiles (DDPs) is an object of interest. 
With a DDP we consider the curve representing the number of people in a defined rectangular region at different time periods during the day. 
We develop a procedure to group similar days in terms of the dynamic of the presences of TIM users considering both the spatial structure and the evolution over the time.
By letting $X_{it}$ be the matrix containing the values of the grid at the quarter $t$ of the day $i$:
\begin{itemize}
\item In the first step, using Histogram of Oriented Gradients (Dalal et al. \cite{dalal2005histogram}, Tomasi \cite{tomasi12histogram}), for each $i$ and for each $t$ we extract the vector of features from $Z_{it}$ (a standardized version of the matrix $X_{it}$ such that min/max is in the interval [0,100] for all $i$ and for all $t$);
\item in the second step, we stack together the vectors of features at a daily basis and we perform a traditional k-means cluster analysis where the objects to be clustered  are the days and the variables are all the stacked vector of features. 
With this step we aim at group days with similarities in the spatial structure of the grid;
\item in the third step, for each cluster, we perform a functional data analysis (FDA) clustering technique (Bouveyron and Come\cite{bouveyron15discriminative}) to further group days. In this step we group days that are similar in terms of the shape of the curve;
\item in the fourth step we define, for each final group, confidence intervals for the DDP, by using functional box plots (Sun and Genton \cite{sun11functional,sun12adjusted}). 
\end{itemize}
Much details can be found in Metulini and Carpita \cite{metulini19tobe}.
The output of the procedure is a functional box plot on the DDPs of a group of similar days. 
For example, by applying the procedure to the rectangular grid defined by latitude 45.516$^{\circ}$ N - 46.564$^{\circ}$ N and longitude 10.18$^{\circ}$ N - 10.245$^{\circ}$ N (roughly corresponding to the municipality of Brescia) and composed by 39 x 39 cells, we find that most of the week days of the Summer 2016 belongs to the same group. Functional box plots of that days (Figure \ref{fig:bplot}) highlight the amount of TIM users along different quarters, that varies, by month and by quarter, from a minimum of 30 to a maximum of about 55 thousands of TIM users.
These values give us a series of information, for example that the number of people in the city increases during the morning and the afternoon hours and decreases during the night. 
However, the same values give us no information on the number of total people. 
\section{The linkage strategy}
\label{sec:strat}
With the procedure in section \ref{sec:proc} we are able to quantify the number of TIM users in the city in selected days and in particular hours of the day.
However, to have an idea of the total number of people we have to consider users of other mobile phone companies as well. 
This data is often unavailable, unless to an onerous cost.
An alternative approach is to apply the TIM mobile phone market share coefficient to the number of TIM users to retrieve the total number of people. 
A country-level estimate being available through ``Il Sole 24 Ore" newspaper.
This value stands to 30.2 \%  (2016, December).
However, we have reasons to think that TIM market share varies along cities since socio-economic and demographic variables (Table \ref{tab:demo}) highlight significant differences.
\begin{table}[htbp!]
		\begin{tabular}{lclc}
			\textbf{Quantity} & \textbf{Municipality of Brescia}	 & &\textbf{Italy}	 \\
			\hline
	\textbf{Per-capita revenues (Euro / year)}$^1$ 		&23,418	& & 19,514	\\
		\textbf{\% foreigners}$^2$	&18.5& &	8.5	\\
		\textbf{Average number of people per family}$^2$	& 2.11 &	& 2.33	\\
		\textbf{Average age}$^2$	&45.8& & 44.7 \\
			\hline	
		1 MEF -Dipartimento delle Finanze (2016) \\
		2 ISTAT (2017)
		\end{tabular}	
		\caption{Socio-economic and demographic comparison between Municipality of Brescia and Italy.}
		\label{tab:demo}
\end{table}
To estimate the market share coefficient for the municipality of Brescia we employ the following strategy. 
We compare the data on the number of residents from administrative archives with the number of TIM users on a selected region during a specific hour of the day.

\begin{figure}[htbp!]
	\sidecaption
	\includegraphics[width=0.6\linewidth]{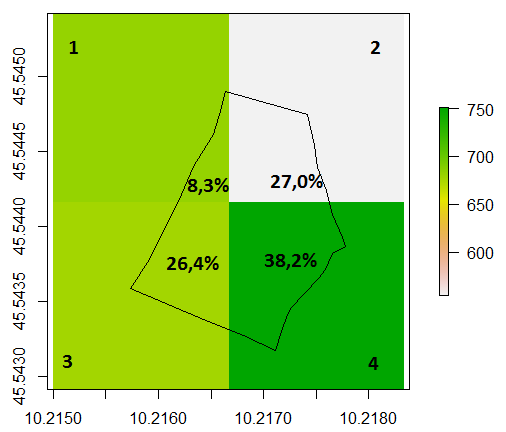}
	\caption{Weighting scheme to assign the number of TIM users to SC 110, located at latitude 45.544$^{\circ}$ N and longitude 10.217$^{\circ}$ N.}
	\label{fig:extract}
\end{figure}

\begin{figure}[!htb]
  	\begin{minipage}{0.47\textwidth}
  		\centering
  		\includegraphics[width=0.55\linewidth]{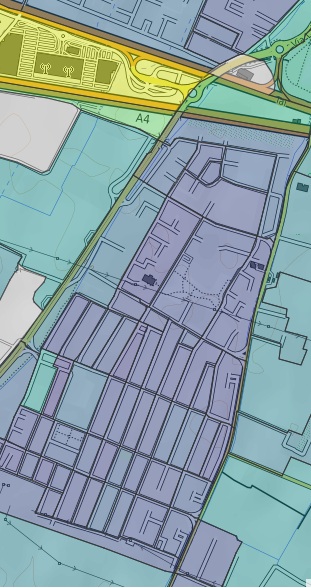}
  		\caption{Zoom of the $ratio$ map for ``Villaggio Sereno" area. Latitude 45.505$^{\circ}$ N - 45.523$^{\circ}$ N and longitude 10.179$^{\circ}$ N - 10.194$^{\circ}$ N.}
  		\label{map:zoom1}
  	\end{minipage}
  	 \hfill
  	\begin{minipage}{0.47\textwidth}
  	\centering
  	\includegraphics[width=0.95\linewidth]{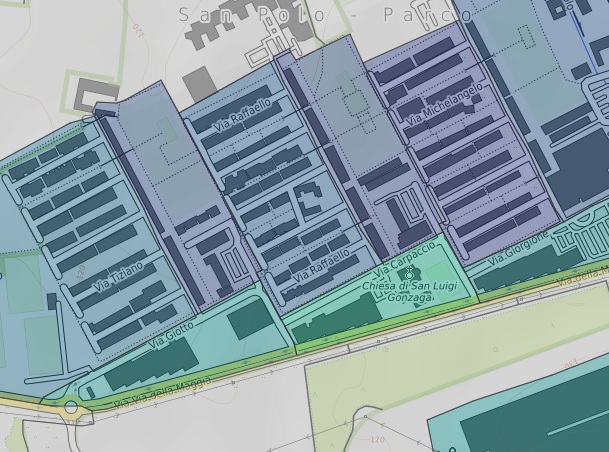}
  	\caption{Zoom of the $ratio$ map for ``San Polo" area. Latitude 45.511$^{\circ}$ N - 45.519$^{\circ}$ N and longitude 10.235$^{\circ}$ N - 10.250$^{\circ}$ N.}
  	\label{map:zoom2}
  \end{minipage}
\end{figure}

To the sake of comparison, we assume that, during late evening hours, residential areas are populated just by residents.
ISTAT (\url{https://www.istat.it/it/archivio/104317}) published geographical data called ``Basi territoriali e variabili censuarie" in the form of a shape file with data (the so-called \texttt{SpatialPolygonDataFrame} in \texttt{R} language).
For the municipalities with more than 20,000 residents, ISTAT aggregates the region at a ``Sezioni di censimento" (SC) level.
To have an idea, the municipality of Brescia has 1,836 SCs.
The shape file contains the information on the number of residents by SC.
We compute the number of TIM users in each SC by putting together the grid data of the TIM users with the shape files containing the value of residents.
The grid cells have regular size while SCs are irregular size polygons.
So, to count the number of TIM users in each polygon we apply a weighted scheme based on the portion of the polygon contained in the cell\footnote{In doing so, we had to convert the coordinate reference system of the spatial polygon object from "UTM" to ``longlat". ISTAT provides shape files in WGS84 international coordinate reference system.}. 
For example, SC 110, located at latitude 45.544$^{\circ}$ N and longitude 10.217$^{\circ}$ N (Figure \ref{fig:extract}) overlaps with 4 cells. Cell 1, at 9pm - October 28th 2015, counts for 682 TIM users, cell 2 counts for 555 users, cells 3 for 677 and cell 4 751. 
Moreover, 8.3\% of the polygon lies in cell 1, 27.0\% in cell 2, 26.4\% in cell 3 and 38.2\% in cell 4.
The number of TIM users in SC 110 is estimated as: 
$$TIM\ users = (682*0.083+555*0.270+677*0.264+751*0.382)* \frac{area(SC)}{area(cell)}$$
After having computed this number for every SC, we then compute the ratio among TIM users and residents\footnote{In the count of ``Residents" we exclude elderly people ($>$80 years) and children ($<$ 11 years), assuming that we want to consider just those with a smartphone.}:
$$Ratio = \frac{TIM\ users}{residents}$$
Some descriptive statistics on the ratio index by SC, evaluated on TIM users at 9pm - October 28th 2015, are reported in Table \ref{tab:quart}.
The median value is consistent with the TIM market share at a country level.
However, the values in the right tail of the distribution are extremely large. For example, the 95th percentile stands to 5.567, meaning that the number of TIM users retrieved in that polygon are five times as large as the number of residents. 
\begin{table}[htbp!]
	\centering
	\begin{tabular}{cccccccccccccc}
	 & \textbf{min}	& & \textbf{5th percentile}   & & \textbf{25th percentile}&	 &\textbf{median}&	&\textbf{75th percentile}&	 & \textbf{95th percentile} & & \textbf{max}	  \\
		\hline
	&	0.006 & &   0.070  & &0.139& & 0.245 & &0.547 & & 5.567 & &347.024 	\\
	\end{tabular}	
	\caption{Quartiles distribution of the $ratio$ index in the municipality of Brescia, by SC. 9pm - October 28th 2015.}
	\label{tab:quart}
\end{table}
\begin{figure}[htbp!]
	\sidecaption
	\includegraphics[width=0.64\linewidth]{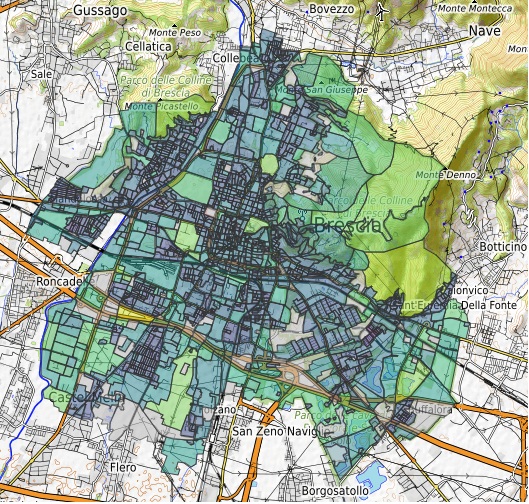}
	\caption{Map of the $ratio$ by SC in the municipality of Brescia, 9pm - October 28th 2015. Colors go from blue (small ratios) to yellow (large ratio).}
	\label{fig:map}
\end{figure}
The map in Figure \ref{fig:map} confirms this evidence\footnote{A dynamic map can be found at authors personal websites.}. 
We find that the ratio value in residential areas is quite homogeneous. Unfortunately, we also notice how TIM assignment of the users in space is affected by the localization of Antennas. 
So, in many areas (particularly those larger ones where the number of residents is small) the $ratio$ index is overestimated. 
This fact suggests that that the Antenna is often localized in large SCs where few people reside. So, people on residential areas are likely tracked where the antenna is (i.e. up to few hundred meters far from where the person is).
For this reason, we decide to visually detect in the map residential areas that are located close to anomalies (i.e. where the antenna is). 
The strategy is to consider the $ratio$ index for a selected residential area plus a portion of region that surrounds the area.
Residential areas are chosen according to DUSAF (Destinazione d'Uso dei Suoli Agricoli e Forestali) maps (  \url{https://www.dati.lombardia.it/Territorio/Dusaf-5-0-Uso-del-suolo-2015/iq6r-u7y2}). 
A zoom of the $ratio$ map is reported in Figure \ref{map:zoom1} (Villaggio Sereno residential areas plus a shopping center) and Figure \ref{map:zoom2} (San Polo residential area plus a factory). The $ratio$ index in the first area stands to 0.265 and the same value stands to 0.310 in the area of San Polo.   
\section{Conclusions}
\label{sec:conc}
Mobile phone data can be used to count the number of people in the city considering the dynamics over the time, differently to administrative archives. 
However, generally, just data on a portion of phone users is available (i.e. just the users of a company).
For this reason we have developed a method to estimate the market share coefficient of the company at a municipality level - whereas the market share is available only at country level - by using administrative data on the number of residents by ``Sezione di censimento".
In doing so we take into account for the bias caused by the location of the antennas.
Results are consistent with the market share at a national level.
   
\begin{acknowledgement}
Authors are grateful with the Statistical Office of the Municipality of Brescia, with a special mention to Dr. Marco Palamenghi, who kindly supported us with providing the data. 
\end{acknowledgement}
%
%
%

\end{document}